# New advances in dual reciprocity and boundary-only RBF methods

W. Chen[*] and M. Tanaka[**]

Department of Mechanical System Engineering, Shinshu University, Wakasato 4-17-1, Nagano City, Nagano 380-8533, Japan (E-mail: [*] chenw@homer.shinshu-u.ac.jp and [**] dtanaka@gipwc.shinshu-u.ac.jp)

This paper made some significant advances in the dual reciprocity and boundary-only RBF techniques. The proposed boundary knot method (BKM) is different from the standard boundary element method in a number of important aspects. Namely, it is truly meshless, exponential convergence, integration-free (of course, no singular integration), boundary-only for general problems, and leads to symmetric matrix under certain conditions (able to be extended to general cases after further modified). The BKM also avoids the artificial boundary in the method of fundamental solution. An amazing finding is that the BKM can formulate linear modeling equations for nonlinear partial differential systems with linear boundary conditions. This merit makes it circumvent all perplexing issues in the iteration solution of nonlinear equations. On the other hand, by analogy with Green's second identity, this paper also presents a general solution RBF (GSR) methodology to construct efficient RBFs in the dual reciprocity and domain-type RBF collocation methods. The GSR approach first establishes an explicit relationship between the BEM and RBF itself on the ground of the weighted residual principle. This paper also discusses the RBF convergence and stability problems within the framework of integral equation theory.

## 1. Introduction

The radial basis function (RBF) is a powerful concept to numerical computations. For example, the introducing the RBF into the BEM [1] has eliminated its major weakness in handling inhomogeneous terms. Nevertheless, the constructing efficient RBFs is still an open research topic. This paper focuses on this problem.

First, we introduce the non-singular general solution as the RBF to derive a meshless, exponential convergence, integration-free, and boundary-only technique. The dual reciprocity method (DRM) and RBF are also employed jointly here to approximate particular solution of inhomogeneous terms as in dual reciprocity BEM (DRBEM) and the method of fundamental solution (MFS). This combined method circumvents either singular integration, slow convergence and mesh in the DRBEM [2] or artificial boundary outside physical domain in the MFS [2,3]. Additionally, the present method holds the symmetric matrix structure for self-adjoint operators subject to a kind of boundary condition. These merits lead to tremendous improvement in computational accuracy, efficiency and stability. The method is named as the boundary knot method (BKM) to differentiate it from the other numerical techniques. The term "boundary-only" is used here in the sense as in the DRBEM and MFS that only boundary knots are required, although internal knots can improve solution accuracy. It is also found that the BKM can produce linear modeling for nonlinear problems with nonlinear governing equations and linear boundary conditions [4]. Three numerical examples are provided to verify the efficiency and utility of this new technique. The completeness concern of the BKM is also discussed.

Second, the RBF has a very close tie with the BEM. By analogy with the BEM weighted residual formulation (also Green's second identity), we present a new general solution RBFs (GSR) construction methodology for inhomogeneous problems in the DRM and the domain-type RBF techniques, which include superconvergent pre-wavelet RBFs.

## 2. Approximation of Particular Solution

The BKM can be viewed as a two-step numerical scheme, namely, approximation of particular solution and the evaluation of homogeneous solution. The former has been well developed in reference [2]. For completeness, we outline its basic methodology. Let us consider the differential equation

$$L\{u\} = f(x) + \rho\{u\} \qquad (1)$$

defined in a region $V$ bounded by a surface $S$, where $L$ represents a differential operator with available non-singular general solution, $\rho\{u\}$ is the remaining differential operator. $f(x)$ is a known forcing function and $x$ means multidimension independent variable. Boundary conditions on $S$ may also be inhomogeneous:

$$u(x) = D(x), \qquad x \subset S_u, \qquad (2a)$$

$$\frac{\partial u(x)}{\partial n} = N(x), \qquad x \subset S_T, \qquad (2b)$$

where $n$ is the unit outward normal. The solution of Eq. (1) can be expressed as

$$u = v + u_p, \qquad (3)$$

where $v$ and $u_p$ are the general and particular solutions, respectively. The latter satisfies equation

$$L_1\{u_p\} = f(x) + \rho\{u\}, \qquad (4)$$

but does not necessarily satisfy boundary conditions. $v$ is the homogeneous solution

$$L\{v\}=0, \qquad (5)$$

$$v(x)=D(x)-u_p, \qquad (6a)$$

$$\frac{\partial v(x)}{\partial n}=N(x)-\frac{\partial u_p(x)}{\partial n}, \qquad (6b)$$

Eqs. (5) and (6a,b) will be later solved by the RBF using non-singular general solution. The DRM analogizes the particular solution by the use of a series of approximate particular solution at all specified nodes of boundary and domain. The inhomogeneous terms of Eq. (4) are approximated by

$$f(x)+\rho(u) \cong \sum_{j=1}^{N+L} \alpha_j \phi(r_j), \qquad (7)$$

where $\alpha_j$ are the unknown coefficients. $N$ and $L$ are respectively the numbers of knots on boundary and domain. $r_j = \|x-x_j\|$ represents the Euclidean norm, $\phi$ is the RBF.

By forcing Eq. (7) to exactly satisfy Eq. (4) at all nodes, we can uniquely determine

$$\alpha = A_\phi^{-1}\left[f(x)+\rho\{A_\phi\}A_\phi^{-1}u\right], \qquad (8)$$

where $A_\phi$ is nonsingular RBF interpolation matrix and $I$ the unite matrix.

Finally, we can get particular solutions at any point by summing localized approximate particular solutions

$$u_p = \sum_{j=1}^{N+L} \alpha_j \varphi(\|x-x_j\|). \qquad (9)$$

Substituting Eq. (8) into Eq. (9) yields

$$u_p = \Phi A_\phi^{-1}\left[f(x)+\rho\{A_\phi\}A_\phi^{-1}u\right], \qquad (10)$$

where $\Phi$ is a known matrix comprised of $\varphi(r_{ij})$. In this study, the approximate particular solution $\varphi$ is determined beforehand, and then we evaluate the corresponding RBF $\phi$ through substituting the specified $\varphi$ into the Helmholtz equation. For the multiquadratic (MQ) RBF, the chosen approximate particular solution is

$$\varphi(r_j)=\left(r_j^2+c_j^2\right)^{3/2}, \qquad (11)$$

where $c_j$ is the shape parameter. The corresponding RBF is

$$\phi(r_j)=6\left(r_j^2+c_j^2\right)+\frac{3r^2}{\sqrt{r_j^2+c_j^2}}+\left(r_j^2+c_j^2\right)^{3/2}, \qquad (12)$$

## 3. Boundary RBF using non-singular general solution

This section deals with the analogization of homogeneous solution by the RBF using non-singular general solution. We take Helmholtz operator as an illustrative example here. There exist non-singular general solutions of the other differential operators [4]. The reason for this choice is that the Helmholtz operator is the simplest among various often-encountered operators having non-singular general solution.

In the standard BEM and MFS, the Hankel function

$$H(r)=J_0(r)+iY_0(r) \qquad (13)$$

is applied as the fundamental solution of the 2D homogeneous Helmholtz equations, where $J_0(r)$ and $Y_0(r)$ are the zero-order Bessel functions of the first and second kinds, respectively. It is noted that $Y_0(r)$ has logarithm singularity, which causes major troubles applying the BEM and MFS.

The present BKM scheme discards the singular general solution $Y_0(r)$ and only use $J_0(r)$ as the radial basis function to collocate the boundary condition equations. Unlike the MFS, all nodes are placed only on physical boundary and can be used as either source or response points. The homogeneous solution of Eq. (5) is collocated by

$$v(x)=\sum_{k=1}^{N} \lambda_k J_0(r_k), \qquad (14)$$

where $r_k = \|x-x_k\|$. $k$ is index of source points on boundary. $N$ is the total number of boundary knots. $\lambda_k$ are the desired coefficients. In terms of the collocation of Eqs. (6a) and (6b), we have

$$\sum_{k=1}^{N} \lambda_k J_0(r_{ik})=D(x_i)-u_p(x_i), \qquad (15)$$

$$\sum_{k=1}^{N} \lambda_k \frac{\partial J_0(r_{jk})}{\partial n}=N(x_j)-\frac{\partial u_p(x_j)}{\partial n}, \qquad (16)$$

where $i$ and $j$ indicate boundary response knots respectively located on $S_u$ and $S_\Gamma$. If internal nodes are used, the following equations at interior knots are supplemented

$$\sum_{k=1}^{N} \lambda_k J_0(r_{lk})=u_l-u_p(x_l), \quad l=1,\ldots L, \qquad (17)$$

where $L$ is the total number of interior points used. Substituting Eq. (10) into Eqs. (15), (16) and (17), we get $N+L$ resulting simultaneous algebraic equations. It is stressed that the use of interior points is not necessary in the BKM. If only boundary knots are employed, Eq. (17) should be omitted.

## 4. Other non-singular general solutions

We here present non-singular general solutions of some often-used 2D and 3D operators. In the following equations $r$ means distance variable and $A$ is coefficients. For 2D equation

$$\nabla^2 u - \lambda^2 u = 0, \qquad (18)$$

the non-singular particular solution is

$$u^* = A I_0(\lambda r), \qquad (19)$$

where $I_0$ is the zero-order modified Bessel function of the first kind. For the 3D Helmholtz-like operator,

$$\nabla^2 u \pm \lambda^2 u = 0, \qquad (20)$$

we respectively have the non-singular general solution

$$u^* = A\frac{\sin(\lambda r)}{\lambda r} \qquad (21)$$

and

$$u^* = A\frac{\sinh(\lambda r)}{\lambda r}, \qquad (22)$$

where sinh denotes the hyperbolic function. For the 2D biharmonic operator

$$\nabla^4 w - \lambda^2 w = 0, \qquad (23)$$

we have the non-singular general solution

$$w^* = A_1 J_0(\lambda r) + A_2 I_0(\lambda r), \qquad (24)$$

where $J_0$ and $I_0$ are respectively the Bessel and modified Bessel functions of the first kind of the zero-order. The non-singular general solution of the 3D biharmonic operator is given by

$$w^* = A_1 \frac{\sin(\lambda r)}{\lambda r} + A_2 \frac{\sinh(\lambda r)}{\lambda r}, \qquad (25)$$

For two-dimensional steady-state convection-diffusion equation

$$D\nabla^2 \phi + v_x \frac{\partial \phi}{\partial x} + v_y \frac{\partial \phi}{\partial y} + k\phi = 0, \qquad (26)$$

where $v_x$ and $v_y$ are the components of the velocity vector $v$, $D$ is the diffusivity coefficient and $k$ represents the reaction coefficient. We have non-singular general solution

$$\phi^* = A e^{\frac{v \cdot r}{2D}} J_0(\mu r), \qquad (27)$$

where

$$\mu = \left[ \left(\frac{|v|}{2D}\right)^2 + \frac{k}{D} \right]^{\frac{1}{2}} \qquad (28)$$

and marked $r$ denotes the distance vector between the source and response knots.

## 5. Numerical Results

The geometry of all test problems is an ellipse with semi-major axis of length 2 and semi-minor axis of length 1 [5]. These examples are chosen since their analytical solutions are obtainable to compare. More complex problems can be handled in the same BKM fashion without any extra difficulty. It is stressed that only boundary knots are employed in these numerical experiments.

### 5.1. Laplace problem

The homogeneous Laplace problem is typically well suited to be solved by the standard BEM technique. In contrast, there appears inhomogeneous term in the BKM formulation to apply the non-singular general solution of the Helmholtz operator. Therefore, this is a persuasive example to show up the accuracy and efficiency of the BKM vis-a-vis the BEM. The Laplace equation is given by

$$\nabla^2 u = 0 \qquad (29)$$

with boundary condition

$$u = x + y. \qquad (30)$$

Eq. (30) is easily found to be a particular solution of Eq. (29). The numerical results are displayed in Table 1, where the BEM solutions come from reference [5].

The MQ shape parameter $c$ is set 25 for both 3 and 5 boundary knots in the BKM. It is observed that the solutions are not sensitive to the parameter $c$. It is surprising to see from Table 1 that the BKM solutions using 3 boundary knots achieve the accuracy of four significant digits and are far more accurate than the standard BEM solution using 16 boundary nodes. This striking accuracy of the BKM with very few knots validates its spectral convergence. In this case only boundary points are employed to approximate the particular solution by the DRM and RBF. It is interesting to note that the coefficient matrix of the BEM and BKM are both fully populated. Unlike the BEM, however, the BKM has symmetric coefficient matrix for all self-adjoint operators with only one type of boundary conditions as this case. We [4] proposed some possible strategy to conserve this symmetric matrix structure in the BKM to the problems with mixed type of the boundary conditions.

Table 1. Results for the Laplace problem

| x | y | Exact | BEM (16) | BKM (3) | BKM (5) |
|---|---|---|---|---|---|
| 1.5 | 0.0 | 1.500 | 1.507 | 1.500 | 1.500 |
| 1.2 | -0.35 | 0.850 | 0.857 | 0.850 | 0.850 |
| 0.6 | -0.45 | 0.150 | 0.154 | 0.150 | 0.150 |
| 0.0 | 0.0 | -0.450 | -0.451 | -0.450 | -0.450 |
| 0.9 | 0.0 | 0.900 | 0.913 | 0.900 | 0.900 |
| 0.3 | 0.0 | 0.300 | 0.304 | 0.300 | 0.300 |
| 0.0 | 0.0 | 0.0 | 0.0 | 0.0 | 0.0 |

### 5.2. Helmholtz problems

To further justify the exponential convergence of the BKM, we consider the inhomogeneous 2D Helmholtz problem governed by equation

$$\nabla^2 u + u = x. \qquad (31)$$

Inhomogeneous boundary condition

$$u = \sin x + x \qquad (32)$$

is posed. It is obvious that Eq. (32) is also a particular solution of Eq. (31). Numerical results by the present BKM is displayed in Table 2.

Table 2. Results for the Helmholtz problem

| x | Y | Exact | BKM (5) | BKM (7) |
|---|---|---|---|---|
| 1.5 | 0.0 | 2.50 | 2.45 | 2.51 |
| 1.2 | -0.35 | 2.13 | 2.08 | 2.14 |
| 0.6 | -0.45 | 1.16 | 1.18 | 1.16 |
| 0.0 | 0.0 | 0.0 | 0.1 | -0.002 |
| 0.9 | 0.0 | 1.68 | 1.66 | 1.69 |
| 0.3 | 0.0 | 0.60 | 0.64 | 0.60 |
| 0.0 | 0.0 | 0.0 | 0.08 | -0.001 |

The numbers in brackets of Table 2 mean the total nodes. The shape parameter $c$ in the MQ is chosen 3. It is found that the present BKM converges very quickly. This shows the BKM holds the super-convergent merit. The BKM can yield accurate solutions with only 7 knots. In contrast, the DRBEM employs16 boundary and 17 interior points to achieve slightly less accurate solutions for a simpler homogeneous case [5]. This is because the normal BEM has only low order of convergence speed [2].

Table 3. Results for Burger equation

| x | Y | Exact | BKM (5) | error % |
|---|---|-------|---------|---------|
| 4.5 | 0.0 | 0.444 | 0.479 | -7.9 |
| 4.2 | -0.35 | 0.476 | 0.515 | -8.2 |
| 3.6 | -0.45 | 0.555 | 0.585 | -5.4 |
| 3.0 | -0.45 | 0.666 | 0.666 | 0.15 |
| 2.4 | -0.45 | 0.833 | 0.808 | 3.1 |
| 1.8 | -0.35 | 1.111 | 1.089 | 2.0 |
| 1.5 | 0.0 | 1.333 | 1.300 | 2.5 |
| 3.9 | 0.0 | 0.512 | 0.563 | -9.7 |
| 3.3 | 0.0 | 0.606 | 0.632 | -4.2 |
| 3.0 | 0.0 | 0.666 | 0.672 | -7.3 |
| 2.7 | 0.0 | 0.740 | 0.725 | 2.0 |
| 2.1 | 0.0 | 0.952 | 0.918 | 3.6 |

### 5.3. Burger equation

An amazing finding of this research is that if only boundary knots are used, the BKM can formulate linear analogization equations for nonlinear problems which have nonlinear governing equations and linear boundary conditions. Consider the Burger equation [5]

$$\nabla^2 u + u_x u = 0. \quad (33)$$

with inhomogeneous boundary condition

$$u = 2/x \quad (34)$$

Eq. (34) is also a particular solution of Eq. (33). Note that the origin of the Cartesian co-ordinates system is dislocated to the node (3,0) to circumvent singularity at $x$=0 as done in [5]. Eq. (33) can be restated as

$$\nabla^2 u + u = u - u_x u. \quad (35)$$

We can use the same procedure as in the previous to solve the above Eq. (35). The shape parameter is taken 1 in the MQ. The resulting BKM formulation is linear algebraic simultaneous equations when only the boundary nodes are used. The BKM results are listed in Table 3 against the analytical solutions. The average relative absolute error is 3.97%. It is worth stressing that the Berger equation has the structure of domain-dominant convection diffusion equation and $u$ in Eq. (35) actually represents velocity. Therefore, it is the convection terms rather than nonlinear constitution which causes the inaccuracy of the solution if not using inner points. Therefore, the use of non-general solution of convection-diffusion equation is expected to perform better in this case. For more detailed discussions see reference [4]. On the other hand, the solution accuracy will also be improved greatly if we use the interior nodes at the expense of sacrificing linear formulation.

### 6. BKM completeness, DRM and domain-type RBFs collocation

Although numerical experiments show that the BKM produced accurate solutions, the possible incompleteness due to only use of the non-singular part of fundamental solution is still a concern. We only numerically tested the interior Helmholtz-type problems. For completeness, we will exam the exterior problems in later experiments. In addition, there still exist some controversies in the choice of basis functions. For example, consider homogeneous biharmonic equation

$$\nabla^4 w = 0, \quad (36)$$

we have four general solutions, namely,

$$w^*(r) = C_1 \ln(r) + C_2 r^2 \ln(r) + C_3 r^2 + C_4. \quad (37)$$

It is common practice in the BEM to use $r^2 \ln(r)$ as the fundamental solution. We are wondering if the other three general solutions will work in the framework of the MFS, BEM or the BKM. Although Duchon [6] proved that $r^2 \ln(r)$ is optimal interpolants for biharmonic operator with linear terms constraints, only use of $r^2 \ln(r)$ is insufficient in the MFS and BKM which require two independent RBFs. Some numerical and theoretical investigations should be carried out. In the MFS, we suggest to use the following two RBFs for completeness:

$$w(r) = A(\ln(r) + 1) + Br^2(\ln(r) + 1), \quad (38)$$

where $A$ and $B$ are undecided coefficients in the RBF.

So far the MQ is only the known RBF with exponential convergence merit. However, the troublesome shape parameter degrades its practical attractiveness. Success applying non-singular general solution in the BKM also hints that it is feasible to develop some operator-dependent RBFs for domain-type collocation methods and the DRM within the BKM, which at least partly satisfy the intrinsic characteristics of the targeted problems. It is expected that operator-dependent RBFs can also achieve superconvergence just as we found in the BKM. The following outlines novel general solution RBFs methodology.

By Green's second theorem (or the weighted residual using fundamental solution) in the indirect BEM, we have solution of Eqs. (1)-(3)

$$u(z) = \int_V G(z,x) f(x) dV(x) + \int_S \left[ u \frac{\partial G(z,x)}{\partial n(x)} - G(z,x) \frac{\partial u(x)}{\partial n(x)} \right] dS(x), \quad (39)$$

where $G$ denotes the fundamental solution of operator $L\{\}$-$\rho\{\}$, $x$ indicates source point. The above formula (39) suggests us that the RBFs can be created for interior source points by

$$\phi(r,x) = [f(x) + \rho(g(r))] r^{-2m} g(r), \quad (40a)$$

where $g(r)$ is the general solution of operator. $x$ represents source point coordinates. $\rho(g(r))$ in formula (40a) may be removed in some cases. $m$ is integral number and $r^{2m}$ term ensures sufficient degree of differential continuity. If we place source and response nodes distinctly to avoid possible singularity, $r^{2m}$ becomes unnecessary in the presented RBFs. Formula (40a) can roughly be interpreted as transforming forcing function $f(x)$ based on eigenfunctions of a operator, which has close relation with the general solution. For Dirichlet and Neumann boundary source points, we respectively have RBFs

$$\phi(r,x) = D(x)r^{2m}\frac{\partial g(r)}{\partial n} \quad (40b)$$

and

$$\phi(r,x) = N(x)r^{2m}g(r). \quad (40c)$$

Normal derivative in Eq. (40b) can be simply replaced by $\partial g(r)/\partial r$. For simplicity, the following RBF is also suggested

$$\phi(r,x) = r^{2m}g(r) \quad (41)$$

for all source points. Thin plate spline (TPS) RBF can be regarded as a special case of Eq. (41). In addition, we can construct pre-wavelet RBFs by substituting $\sqrt{r_j^2 + c_j^2}$ into Eqs. (40a,b,c) and (41) instead of $r$, where $c_j$ is dilution parameters. For example, numerical experiments using $r_j^{2m}\ln\sqrt{r_j^2 + c_j^2}$ manifest spectral convergence as in the MQ. Such wavelet RBFs will be especially attractive for multiscale problems. The modified TPS (MTPS) RBF $r^{2m}(ln(r)+1)$ is suggested based on the consideration of the general solutions of the Laplace and biharmonic operators.

The RBF representation is stated as

$$u(x_i) = \sum_{k=1}^{N} \beta_k \phi(r_{ik}, x_k) + \beta_{N+1}\psi(x_i), \quad (42)$$

where $\psi(x)$ is the linear constraint. The side condition is

$$\sum_{k=1}^{N} \beta_k \psi(x_k) = 0. \quad (43)$$

If the non-singular solution is used in Eqs. (40a-c) without $r^{2m}$ term, the linear constraint is not necessary.

Although in recent years great effort has been devoted to developing a formal mathematical theory of the RBF, some very limited advances are achieved by using a sophisticated interpolation analysis. There are still many essential issues unsolved such as convergence, slovability, stability and construction of proper RBF. The present work strongly suggests that the RBF technique is closely related to integral equation theory such as the Fredholm and Volterra integral equations. The kernel function of the integral equations is considered the key to create very efficient RBFs for a broad range of problems such as numerical solution of PDE, network, optimization, data processing, and inverse problems. Accordingly the currently popular TPS and MQ are not recommended in many cases. Also the convergence, stability and slovability of the RBF interpolation matrix can readily be established within the theoretical framework of the integral equation theory. More details on these aspects will be provided in the sequent paper.

The present study shows that the RBF technique has very close relationship with the BEM on the ground of the weighted residuals method. Therefore, one can easily conclude that the constructing efficient RBFs using the general solution of the operator is not restricted to the potential problems.

## 7. Conclusions

The present BKM can be considered one kind of the Trefftz method [8] with features of the RBF using non-singular general solution. The method shows that singularity is not an essential ingredient in the boundary-only techniques. Comparing to the indirect BEM, this note presents general solution RBFs for various inhomogeneous problems in the domain-type RBF and DRM approximations, which fully exploit features of certain problems. Golberg and Chen [2] put the DRM on the solid RBF theory. The present work further establishes direct relationship between the RBF itself and the BEM. It is stressed that the RBF schemes may be especially attractive for higher dimension problems due to their dimensionally increased order of convergent rate [7], which makes it circumvent dimension curse.

**Acknowledgements**
The authors express grateful acknowledgments of helpful discussions with Profs. C.S. Chen, M. Golberg. Y.C. Hon, and J.H. He.